\documentclass[sigplan, preprint, noacm=true, 10pt]{acmart}

\AtBeginDocument{%
  \providecommand\BibTeX{{%
    Bib\TeX}}}

\setcopyright{none}
\renewcommand\footnotetextcopyrightpermission[1]{}

\usepackage[utf8]{inputenc}
\usepackage[T1]{fontenc}
\usepackage[abbreviations]{foreign}
\usepackage{graphicx}
\usepackage{subcaption}
\usepackage[margin=3pt,skip=6pt,belowskip=0pt,font={small,stretch=0.9},labelfont=bf]{caption}
\usepackage{microtype}
\usepackage{setspace}
\usepackage{listings}
\usepackage{xcolor}
\usepackage{tabularx}
\usepackage{multirow}
\usepackage{refcount}
\usepackage{fp}

\lstdefinelanguage{JavaScript}{
	morekeywords=[1]{break, continue, delete, else, for, function, if, in,
		new, return, this, typeof, var, void, while, with},
	morekeywords=[2]{false, null, true, boolean, number, undefined,
		Array, Boolean, Date, Math, Number, String, Object},
	morekeywords=[3]{eval, parseInt, parseFloat, escape, unescape},
	sensitive,
	morecomment=[s]{/*}{*/},
	morecomment=[l]//,
	morecomment=[s]{/**}{*/}, 
	morestring=[b]',
	morestring=[b]"
}[keywords, comments, strings]
\usepackage{textcomp}
\usepackage{booktabs}
\usepackage{xspace}
\usepackage{url}
\usepackage{pifont}
\usepackage[inline]{enumitem}

\setlist{noitemsep,topsep=0pt,parsep=0pt,partopsep=0pt}

\usepackage{algorithm}
\usepackage{algpseudocode}
\usepackage{hyperref}
\usepackage[capitalise]{cleveref}

\usepackage{tikz}

\definecolor{lightcolor}{rgb}{0,0.5,1}

\usepackage{setspace}
\usepackage[margin=5pt,font={stretch=0.9}]{caption}
\usepackage{fontawesome5}
\newboolean{showcomments}
\setboolean{showcomments}{true}
\ifthenelse{\boolean{showcomments}}
{ \newcommand{\mynote}[3]{
		\fbox{\bfseries\sffamily\scriptsize#1}
		{\small$\blacktriangleright$\textsf{\emph{\color{#3}{#2}}}$\blacktriangleleft$}}}
{ \newcommand{\mynote}[3]{}}

\definecolor{darkgreen}{rgb}{0.3,0.5,0.3}
\definecolor{darkblue}{rgb}{0.3,0.3,0.5}
\definecolor{darkred}{rgb}{0.5,0.3,0.3}

\newcommand{\sys}{\textsc{SecV}\xspace}

\newcommand{\graal}{\textsc{GraalVM}\xspace}
\newcommand{\truffle}{\textsc{Truffle}\xspace}
\newcommand{\polytaint}{\textsc{PolyTaint}\xspace}

\newcommand{\js}{JS\xspace}
\newcommand{\python}{Python\xspace}

\newcommand{\secv}{\textsc{SecV}\xspace}

\newcommand{\code}[1]{\texttt{\small #1}\xspace}

\newcommand{\ecall}{\texttt{\small ecall}\xspace}
\newcommand{\ocall}{\texttt{\small ocall}\xspace}
\newcommand{\ecalls}{\texttt{\small ecall}s\xspace}
\newcommand{\ocalls}{\texttt{\small ocall}s\xspace}

\newcommand{\imagespace}{-3mm}

\definecolor{codegreen}{rgb}{0,0.6,0}
\definecolor{codegray}{rgb}{0.5,0.5,0.5}
\definecolor{codepurple}{rgb}{0.58,0,0.82}
\definecolor{backcolour}{rgb}{0.95,0.95,0.92}

\newcounter{numobserv} 
\definecolor{beaublue}{rgb}{0.9, 0.95, 0.95}
\usepackage{tcolorbox}
\colorlet{shadecolor}{beaublue}
\tcbset{
	colback=beaublue,
	boxrule=0.5pt,
	boxsep=1pt,
	left=1pt,
	right=1pt,
	top=1pt,
	bottom=1pt,
	sharp corners,
	colframe=white,
}
\newcommand{\observ}[1]{
	\addtocounter{numobserv}{1}
	\begin{tcolorbox}	
		\emph{\textbf{Take-away\,\thenumobserv\,:} #1 }	
	\end{tcolorbox}
}

\usepackage{tikz}

\newcommand{\copyrighttext}{ \scriptsize \textcopyright 2023 ACM.               
	Personal use of this material is permitted.                                 
	Permission from ACM must be obtained for all other uses,                   
	in any current or future media, including reprinting/republishing this      
	material for advertising or promotional purposes, creating new collective   
	works, for resale or redistribution to servers or                           
	lists, or reuse of any copyrighted component of this work in other works.   
	This is the author’s version of the work. The final version is published in the proceedings of the 24th International Middleware Conference. \\DOI: \href{https://doi.org/10.1145/3590140.3629116}{10.1145/3590140.3629116}}

\setcopyright{none}
\begin{document}

\title{SecV: Secure Code Partitioning via Multi-Language Secure Values}


\author{Peterson Yuhala}
\affiliation{%
	\institution{University of Neuchâtel}
	\city{Neuchâtel}
	\country{Switzerland}
}
\email{peterson.yuhala@unine.ch}

\author{Pascal Felber}
\affiliation{%
	\institution{University of Neuchâtel}
	\city{Neuchâtel}
	\country{Switzerland}
}
\email{pascal.felber@unine.ch}

\author{Hugo Guiroux}
\affiliation{%
	\institution{Oracle Labs}
	\city{Z\"urich}
	\country{Switzerland}	
}
\email{hugo.guiroux@oracle.com}

\author{Jean-Pierre Lozi}
\affiliation{%
	\institution{Inria}
	\city{Paris}
	\country{France}	
}
\email{jean-pierre.lozi@inria.fr}

\author{Alain Tchana}
\affiliation{%
	\institution{Grenoble INP}
	\city{Grenoble}
	\country{France}	
}
\email{alain.tchana@grenoble-inp.fr}

\author{Valerio Schiavoni}
\affiliation{%
	\institution{University of Neuchâtel}
	\city{Neuchâtel}
	\country{Switzerland}	
}
\email{valerio.schiavoni@unine.ch}

\author{Gaël Thomas}
\affiliation{%
	\institution{Télécom SudParis}
	\city{Institut Polytechnique de Paris}
	\country{France}	
}
\email{gael.thomas@telecom-sudparis.eu}

\renewcommand{\shortauthors}{Yuhala, et al.}

\ccsdesc[500]{Security and privacy~Trusted computing}
\ccsdesc[500]{Software and its engineering~Object oriented frameworks}

\keywords{Trusted Execution Environments, Intel SGX, Managed Execution Environments, Java, Truffle, GraalVM}

\newcommand{\copyrightnotice}{\begin{tikzpicture}[remember picture,overlay]       
	\node[anchor=south,yshift=2pt,fill=yellow!20] at (current page.south) {\fbox{\parbox{\dimexpr\textwidth-\fboxsep-\fboxrule\relax}{\copyrighttext}}};
	\end{tikzpicture}
}

\begin{abstract} 
Trusted execution environments like Intel SGX provide \emph{enclaves}, which offer strong security guarantees for applications.
Running entire applications inside enclaves is possible, but this approach leads to a large trusted computing base (TCB).
As such, various tools have been developed to partition programs written in languages such as C or Java into \emph{trusted} and \emph{untrusted} parts, which are run in and out of enclaves respectively.
However, those tools depend on language-specific taint-analysis and partitioning techniques.
They cannot be reused for other languages and there is thus a need for tools that transcend this language barrier.

We address this challenge by proposing a multi-language technique to specify sensitive code or data, as well as a multi-language tool to analyse and partition the resulting programs for trusted execution environments like Intel SGX.
We leverage GraalVM's \truffle framework, which provides a language-agnostic abstract syntax tree (AST) representation for programs, to provide special AST nodes called \emph{secure nodes} that encapsulate sensitive program information.
Secure nodes can easily be embedded into the ASTs of a wide range of languages via \truffle's \emph{polyglot API}.
Our technique includes a multi-language dynamic taint tracking tool to analyse and partition applications based on our generic secure nodes.
Our extensive evaluation with micro- and macro-benchmarks shows that we can use our technique for two languages (Javascript and \python), and that partitioned programs can obtain up to $14.5\%$ performance improvement as compared to unpartitioned versions.
\end{abstract}


\maketitle
\copyrightnotice
\pagestyle{plain}

\section{Introduction}
\label{sec:introduction}

Trusted execution environments (TEE) use hardware cryptography to enforce confidentiality, authenticity and integrity of a memory zone called an \emph{enclave}. 
TEEs are at the basis of confidential computing, and major CPU vendors have introduced them in their processors, \eg Intel's SGX \cite{vcostan}, AMD's SME \cite{amd}, and ARM's TrustZone \cite{pinto2019demystifying,amacher2019performance}, to provide security guarantees for cloud-based applications. 

Minimising the trusted computing base (TCB) of a TEE-enabled program is crucial for improving security.
Achieving this goal is especially difficult for managed languages because they come with a runtime that contains large system libraries (\eg Java, \python, R, \etc).
As such, leveraging tools such as SCONE \cite{scone}, SGX-LKL \cite{sgxlkl}, or Graphene-SGX \cite{graphene} that fully embed the language runtime inside the TEE is not satisfactory.
For example, the Java library shipped with OpenJDK18 contains 47{,}146 Java classes for a total of 5{,}224{,}426 lines of Java code.
Adding such a large code base inside a TEE increases the size and the attack surface of the TCB to an unacceptable degree.

To minimise the TCB size of applications written in managed languages, the code and the data of the application and the runtime must be partitioned into \emph{trusted} parts and \emph{untrusted} parts, which run inside and outside of the enclave, respectively.
Since manually partitioning a large code base is complex and error-prone, several middleware technologies have been developed which permit developers to automatically partition their code prior to deployment in the cloud.
Unfortunately, partitioning tools only exist for a few languages: Java with Civet \cite{civet}, Montsalvat \cite{montsalvat} and Uranus \cite{uranus}, Go with GOTEE \cite{ghosn2019secured}, and C with Glamdring \cite{glamdring}.
These tools depend heavily on the language semantics and cannot be reused for other languages.
To support the next programming language (\eg \python, R, JavaScript, Ruby, \etc), one must entirely re-implement a new automatic partitioning tool, which is time consuming and error-prone.

In this paper, we propose \emph{to decouple the partitioning tool from the language semantics, and to implement the former once and for all languages.} 
To that end, we leverage  \truffle \cite{truffle}, a Java library for building high-performance language interpreters.
It provides a generic abstract syntax tree (AST) composed of nodes that represent various syntactic elements of a program: \ie expressions (\eg function calls or arithmetic operations), program values (\eg literals or variables), control flow (\eg if-else or for loops), \etc. \truffle provides support for popular programming languages like Python, JavaScript, R, Ruby, C/C++, amongst many others.
To implement the partitioning tool once and for all, we first introduce a new multi-language AST node type, called a \emph{secure node}.
This node contains a \emph{secure value} corresponding to a sensitive value that has to be secured in the enclave.
Then, we leverage the \emph{polyglot interoperability protocol} provided by \truffle \cite{grimmer2015}, which allows the declaration of a secure value from any language with an expression as simple as \code{x\,=\,polyglot.eval("secV",\,"sInt(42)")}.
Finally, we develop a generic \emph{dynamic taint tracking} tool, \polytaint, which instruments \truffle ASTs to track the data flow of sensitive values from secure nodes so as to determine the portions of a program (\eg \emph{functions}) to be shielded inside the enclave. 
\polytaint then partitions the program into two parts, trusted and untrusted, to be executed respectively inside or outside the enclave.



In summary, we propose the following contributions:
\begin{enumerate}[noitemsep]
	\item Generic AST secure nodes to specify sensitive data in any \truffle language.
	\item \polytaint, a \truffle instrumentation tool which performs dynamic taint tracking on generic polyglot programs and partitions the programs based on the use of secure values.
	\item An extensive experimental evaluation demonstrating the effectiveness of our approach via micro-benchmarks in JavaScript and \python, as well as real-world applications: \emph{PageRank} \cite{pagerank} and \emph{linear regression} \cite{regressionXu2022}. Our analysis of partitioned programs shows we can reduce the size of the TCB (\ie improved security) and improve performance (up to $14.5\%$) at the same time.

\end{enumerate}

The rest of the paper is organised as follows. 
\S\ref{sec:background} provides background concepts.
We present our threat model in \S\ref{sec:threatmodel}. 
\S\ref{sec:contributions} describes the architecture and workflow of \sys, followed in \S\ref{sec:evaluation} by an extensive experimental evaluation.
We discuss the limitations of \sys in \S\ref{sec:limitations}, and provide ideas for further use cases in \S\ref{sec:discussions}. Related work is discussed in \S\ref{sec:rw}, while we conclude and hint at future work in \S\ref{sec:conclusion}.

\section{Background}
\label{sec:background}

\subsection{Intel software guard extensions (SGX)}
\label{intelsgx}

Intel SGX \cite{vcostan,sgx2Xing2016} is a set of hardware instructions to create secure memory regions called \emph{enclaves}. 
Enclaves provide strong confidentiality and integrity guarantees for code and data processed on a malicious node whose privileged software (\eg kernel, hypervisor, \etc) is potentially compromised.
Enclave data is stored in an encrypted portion of DRAM, the \emph{enclave page cache} (EPC). 
EPC pages are transparently decrypted by a memory encryption engine (MEE) only when loaded into a CPU cache line. 
Enclaves typically have limited memory resources: 128\;MB or 256\;MB per socket in the more popular first generation SGX-enabled CPUs. 
The SGX Linux kernel driver supports paging from EPC memory to regular DRAM to accommodate enclaves larger than the EPC, but this comes at a performance cost \cite{sgxperf}.

Software that leverages Intel SGX is usually split in two parts: a \emph{trusted} part which executes in enclave mode, and an \emph{untrusted} part which executes in non-enclave mode.
The Intel SGX SDK permits interaction between both parts via \ecalls (from untrusted to trusted code), and \ocalls (from trusted to untrusted).  
Intel SGX enclaves operate only in user mode and OS services (\ie system calls) cannot be executed directly within the enclaves \cite{graphene,scone}.
They must instead be relayed to the untrusted part via \ocalls. 
Both \ecalls and \ocalls trigger expensive context switches in the CPU \cite{sgxperf} (accounting for up to 13{,}500 CPU cycles). 
Reducing the number of \ecalls and \ocalls is therefore key in designing efficient enclave software.

To facilitate the deployment of code in enclaves, several tools \cite{scone, graphene,occlum, sgxlkl} make it possible to run unmodified applications inside SGX enclaves. 
While this approach is relatively straightforward to use for developers, it increases the chance of introducing security vulnerabilities into the enclave (due to the large TCB). 
Some language-specific alternatives \cite{glamdring,civet,montsalvat,uranus} have been proposed recently to partition programs and keep only sensitive code and data within enclaves.

\subsection{GraalVM}
\label{graalvm}

GraalVM is a high-performance JDK distribution that can execute programs implemented in a wide range of high-level languages, \eg JavaScript, \python, Java, Kotlin, \etc \cite{bonettagraalvm,montsalvat}. 
At the heart of GraalVM is the \emph{Graal} compiler \cite{graalcompiler}, a dynamic just-in-time (JIT) compiler that produces highly optimised machine code.

\subsubsection*{GraalVM native images}
\label{native-image}

GraalVM provides a tool that can compile ahead-of-time (AOT) programs implemented in JVM-based languages (\eg Java, Scala, \etc) to native executables, called \emph{native images}.
This tool leverages static analysis \cite{nativeImgs} to determine which program elements (classes, methods, fields) are reachable at run time. 
These reachable elements, together with runtime components (\ie the garbage collector, support for thread scheduling and synchronisation, \etc) are AOT-compiled into the final native image. 
This significantly lowers the memory footprint when compared to programs interpreted in a regular JVM, making native images very suitable for TEEs. 
In addition, GraalVM native images can run initialisation code at build time instead of at run time \cite{nativeImgs}, which leads to faster start-up times in native images compared to other runtime environments.

\subsubsection*{\truffle}
\label{truffle_framework}

\begin{figure}[!t]
	\centering
	\includegraphics[scale=0.8]{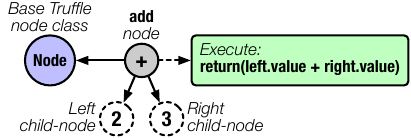}
	\caption{Integer addition node executed in the AST to produce the sum of the values of left and right child.}
	\label{fig:add-node}
\end{figure}

It is a framework provided by the GraalVM ecosystem to build tools and programming language implementations as self-modifying Abstract Syntax Tree (AST) interpreters \cite{trufflesite,grimmer2015}.
At low level, \truffle provides a base \emph{Node} class that is leveraged by language implementers to build other AST nodes representing the semantic constructs of their programming language, \eg an addition operation, a variable write, \etc. 
Essentially, every node in a \truffle AST is executed (via an \code{execute} method) at runtime to produce a result. 

In \autoref{fig:add-node} for example, the integer addition node will be executed at runtime to produce the sum of the left and child node values, which are computed by calling the \emph{execute} methods of the left and right nodes respectively (depth-first traversal).


The key advantage of \truffle is that all programs implemented in a supported language are parsed to a common AST representation (\ie the \truffle AST), which is then manipulated in a language-agnostic fashion.
\truffle languages include JavaScript (\js) \cite{trufflejs}, Ruby \cite{truffleruby}, R \cite{fastr}, \python \cite{graalpython}, LLVM-based languages \cite{sulong}, and more.

\paragraph*{Polyglot API}
\truffle allows developers to build \emph{polyglot} applications that combine code written in different languages. 
This interoperability is provided by the \emph{polyglot interoperability protocol}, a set of standardised messages (\emph{polyglot API}) implemented in every \truffle language \cite{grimmer2015,polyprog}. This API allows to transfer objects from one language scope to another as \truffle \emph{values}, \ie an instance of \code{Value} class.


\begin{lstlisting}[language=Java,
xleftmargin=25pt,
xrightmargin=25pt,
label={lst:polyjavajs},
caption=Java application accessing an object from a \js language scope via the polyglot API.]
import org.graalvm.polyglot.*;
class PolyglotTest {
	public static void main(String[] args) {
		Context polyglot = Context.create();
		Value array = polyglot.eval("js", "[1, 2, 3, 44]");
		int result = array.getArrayElement(3).asInt();
		System.out.println(result);  // prints 44 
	}
}
\end{lstlisting}

\autoref{lst:polyjavajs} shows the transfer of an array object created in a \js language scope into the Java language scope via the polyglot API (line 6). The first parameter to the \code{polyglot.eval} call is the truffle \emph{language id}, a unique string identifier for \truffle guest languages, while the second parameter is the guest source code to execute. Such cross-language object exchange is possible in all \truffle languages, which simplifies the implementation of multi-language applications.

In this paper, the language accessing the objects generated by another interpreter will be referred to as the \emph{host language} (Java in \autoref{lst:polyjavajs}), and the language used to generate the objects accessed by the host language is called the \emph{guest language} (\js in \autoref{lst:polyjavajs}). 
Objects of the host language will be referred to as \emph{regular objects} while those of the guest language will be referred to as \emph{foreign objects}. 

\begin{figure}[!t]
	\centering
	\includegraphics[scale=0.8]{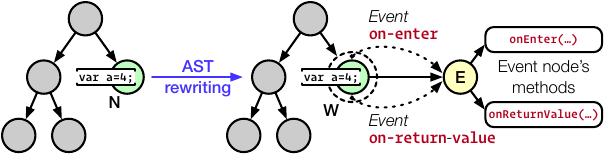}
	\caption{Instrumenting a variable write with a wrapper node. The event node's \code{onReturnValue} method could be used to register the name of the variable being written, as well as the corresponding value.}
	\label{fig:var-write-wrapper}
\end{figure}

\paragraph*{\truffle instrumentation agents}
\label{instrumentation_agents}
The \truffle framework provides an API to dynamically intercept the execution of nodes in the AST \cite{van2018fast, grimmer2015, wurthinger2012}.
This API was used to implement a program profiler \cite{graalprof}, a debugger \cite{chromedebugger}, and a taint-tracking tool \cite{kreindl20}.
In \polytaint, we leverage this API to implement our partitioning tool.

To intercept the execution of nodes, the developer has to implement an \emph{instrumentation agent} \cite{van2018fast}.
The agent intercepts the execution of nodes by leveraging \emph{syntactic tags} associated to the AST nodes.
These tags give the semantics of the nodes (\eg \emph{call} tag, \emph{variable write} tag \etc).

\autoref{fig:var-write-wrapper} illustrates how an instrumentation agent works.
When the agent is loaded, it associates an \emph{event node} to a tag \cite{eventnode}.
In our example, the agent attaches the event node $E$ to any AST node with the \emph{variable write} tag.
At execution, when \truffle visits an AST node with the instrumented tag for the first time, it replaces the node by a wrapper node \cite{wrappernode} connected to the event node.
In our example, when \truffle visits the node $N$ for the first time, it wraps $N$ in $W$ and connects $W$ with $E$.

A wrapper node implements a special \code{execute} function.
Upon a call, it first calls \code{onEnter} on the associated event node, then \code{execute} on the wrapped node, and finally \\\code{onReturnValue} on the associated event node.
These functions can collect metrics, modify the wrapped node, and even replace the wrapped node by another node.

\section{Threat model}
\label{sec:threatmodel}

We consider a powerful adversary with full control over the software stack, including privileged software (\ie host OS, hypervisor) with access to the physical hardware (\ie DRAM, secondary storage, \etc). 
The adversary's goal is to disclose sensitive data or damage its integrity.

The \sys workflow assumes enclave program development, taint tracking and program partitioning, as well as final enclave code building and signing are all done in a trusted environment, to prevent malicious code tampering disclosing sensitive information at runtime. 
The integrity of the trusted partition can be ensured via remote attestation~\cite{vcostan,opera}. 

The adversary cannot open the CPU package to extract decrypted enclave secrets. 
The final enclave code does not intentionally leak sensitive information (\eg encryption keys). 
We do not consider denial-of-service (DoS) and side-channel attacks~\cite{van2018foreshadow, schwarz2017malware}, for which mitigations exist~\cite{gruss2017strong,oleksenko2018varys}.

\section{Design and workflow of \sys}
\label{sec:contributions}


\begin{figure}[!t]
	\centering
	\includegraphics[scale=0.8]{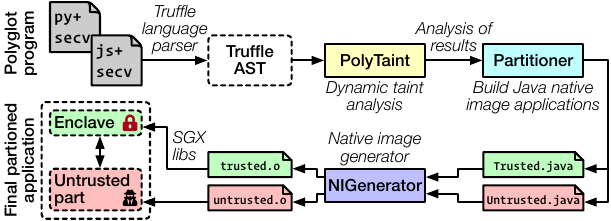}
	\caption{Overview of \sys's workflow.}
	\label{fig:arch}
	\vspace{\imagespace}
\end{figure}

\sys is a framework that analyses applications and partitions them into \emph{trusted} and \emph{untrusted} parts for architectures with TEE support. It supports applications written in \emph{any} Truffle language, and introduces \emph{secure values} to specify sensitive data.
\autoref{fig:arch} presents an overview of \sys's workflow, which comprises 4 phases:
\emph{(1)}~identification of sensitive variables that contain \emph{secure values} (see \S\ref{sec:data_specification}),
\emph{(2)}~dynamic taint tracking with \polytaint (see \S\ref{sec:taint_tracking} and \S\ref{sec:ast-instrumentation}),
\emph{(3)}~program partitioning with the \emph{partitioner} (see \S\ref{sec:partitioning}), and 
\emph{(4)}~native image building with the \emph{image generator} and final application creation (see \S\ref{sec:ImageGenerator}).

We illustrate our system’s workflow by considering a simple linear regression program written in \python (see \autoref{lst:python-regression}). 
We consider a security scenario where we wish to keep the learned model (\ie \code{m} and \code{c}) confidential~\cite{intel-sgx-tf,secureTf,plinius,secureml}.
In the following, we will show how one can use \sys to enforce this scenario.

\begin{figure}[!h]
	\noindent\begin{minipage}[t]{.45\columnwidth}
		\begin{lstlisting}[language=Python,caption={},label={lst:regression},lastline=17,numbersep=-6pt]
		import polyglot
		m = 0.0  (*@\label{lst:python-regression:m}@*)
		c = 0.0 (*@\label{lst:python-regression:c}@*)
		L = 0.0001  
		N = 10000  
		numIter = 100 (*@\vspace{5pt}@*)  
		def readXData(n):  
			# logic (omitted)
			return xData(*@\vspace{5pt}@*)
		def readYData(n): 
			# logic (omitted)
			return yData(*@\vspace{5pt}@*)
		def arraySum(A):  
			sumA = 0
			for a in A:
				sumA += a
			return sumA
		\end{lstlisting}
	\end{minipage}\hfill
	\begin{minipage}[t]{.43\columnwidth}
	\begin{lstlisting}[language=Python,caption={},label={lst:train-model},firstnumber=18,
	lastline=27,xleftmargin=-50pt, numbersep=-10pt]
		def trainModel(numIterations):  
			X = readXData(N)
			Y = readYData(N)
			for i in range(numIterations):
				Y_pred = m * X + c 
				D_m = (-2/N) * arraySum(X * (Y-Y_pred))
				D_c = (-2/N) * arraySum(Y-Y_pred)(*@\label{lst:python-regression:array}@*)
				m = m - L * D_m
				c = c - L * D_c(*@\vspace{5pt}@*)
		trainModel(numIter)  		
	\end{lstlisting}
	\end{minipage}	
	\begin{lstlisting}[caption={Illustrative example: simple linear regression program written in \python.},numbers=none, label={lst:python-regression} ]
	\end{lstlisting}
	\vspace{\imagespace}	
\end{figure}


\subsection{Identifying and specifying sensitive data}
\label{sec:data_specification}	
\begin{figure}[!t]
	\centering
	\includegraphics[scale=0.8]{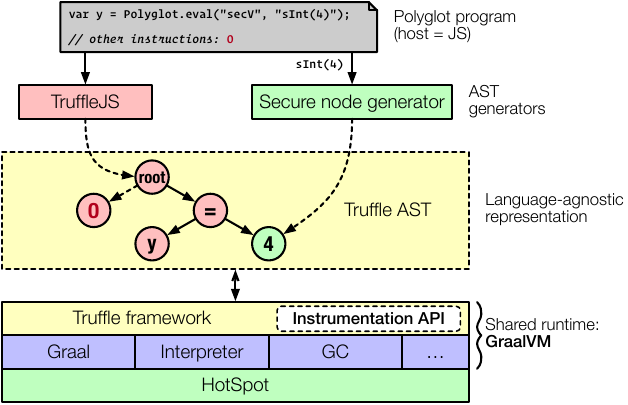}
	\caption{The \truffle framework can be leveraged to provide an AST generator that produces secure nodes which can be injected in any \truffle AST via the polyglot API.}
	\label{fig:secure-value-generator}
	\vspace{\imagespace}
\end{figure}

%
We propose a multi-language approach to specify sensitive data via the use of newly-introduced \emph{secure nodes} in a \truffle AST.
All values associated with these secure nodes are referred to as \emph{secure values} and they will represent sensitive information which must be kept inside the enclave.
To make this possible, we leverage the \truffle framework to build a \emph{secure node generator} which injects secure nodes into any \truffle AST via the polyglot API.
This secure node generator is implemented like a regular \truffle language but provides primarily special AST nodes which comprise secure information.
Our preliminary prototype provides secure node types: \code{sInt}, \code{sDouble}, \code{sBoolean}, and \code{sArray} to contain secure int values, secure double values, secure boolean values, and secure array object values respectively.
For example, \autoref{lst:js-secure-int} shows how a JavaScript program can inject a secure integer value into its AST. 

\begin{lstlisting}[language=JavaScript,
xleftmargin=25pt,
xrightmargin=25pt,
label={lst:js-secure-int},
caption= Injecting a secure integer node with value 4 into a JavaScript program via the polyglot API.]
var myInt = 2;
var secInt = Polyglot.eval("secV", "sInt(4)");
myInt = secInt + 2;
console.log(myInt); // prints 6
\end{lstlisting}


Going back to our illustrative example, since our goal is to secure our ML model (represented by \code{m} and \code{c}), we can specify \code{m} and \code{c} as secure values using the \truffle polyglot API for \python.
To do this, we change the variable assignments for \code{m} and \code{c} to the code shown in \autoref{lst:secure-mc} (lines~\ref{lst:python-regression:m} and~\ref{lst:python-regression:c}).

\FPeval{\firstline}{\getrefnumber{lst:python-regression:m}}
\begin{lstlisting}[language=Python,
xleftmargin=25pt,
xrightmargin=25pt,
label={lst:secure-mc},
firstnumber=\firstline,
caption=Specifying m and c as secure values using \sys.]
m = polyglot.eval(language="secV", string="sDouble(0.0)") 
c = polyglot.eval(language="secV", string="sDouble(0.0)")  

\end{lstlisting}

At run time, the \truffle AST corresponding to the \python program will have \emph{SecV} nodes associated with \code{m} and \code{c} once the \code{polyglot.eval} call is executed.
\sys then analyses the program's AST to determine which portions of the program access the secure values \code{m} and \code{c}.

\smallskip\noindent\textbf{Support for other secure node types.}
\label{sec:new-types}
Implementing more secure types, \eg lists, maps, \etc in \secv is straightforward. It involves: defining a new \truffle node for the type, implementing its internal representation, defining type operations and conversions, and integrating the new type node into \secv's language grammar. To simplify the development of complex types, \truffle's design allows the use of existing Java types, \eg arrays, sets, \etc as building blocks when defining the new node.

\subsection{Taint tracking}
\label{sec:taint_tracking}

\sys includes \polytaint, a \truffle code instrumentation tool that supports applications written in any \truffle language.
\polytaint is designed as a taint tracking tool.
It monitors the creation of secure nodes in a polyglot application and marks a node that uses a value from a secure node (\ie secure value) as secure itself.

\smallskip\noindent\textbf{Assumptions.}
Our preliminary implementation of \sys assumes a \emph{procedural program structure} where the application to be partitioned is organized as a group of $n$ functions: $f_1$, $f_2$, $\ldots$, $f_n$.
Most programming languages support this paradigm.
The goal of \sys is thus to determine the set of program functions to be put inside the enclave, and those to be kept out of the enclave, and partition the program into two parts based on this information. We assume that the inputs used in dynamic analysis are sufficiently exhaustive to cover all legitimate production configurations; this mitigates leakage of sensitive data at runtime due to incomplete code coverage during analysis.

\subsubsection{Taint propagation} 

\begin{figure}[!t]
	\centering
	\includegraphics[scale=0.8]{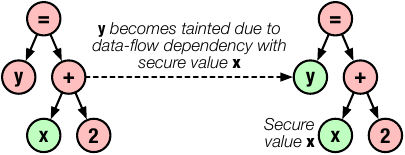}
	\caption{Taint propagation due to explicit data-flow dependencies with secure values.}
	\label{fig:taint-propagation}
\end{figure}

In the context of our work, \emph{taint propagation} defines how secure values flow from the secure nodes into subsequent parts of the program.
\polytaint performs taint propagation via \emph{explicit information flow} \cite{dytan}.
That is, an untainted variable \code{y} becomes tainted if an already tainted variable \code{x} is directly involved in the computation of \code{y}'s value.
In other words, there is a direct data-flow dependency between \code{x} and \code{y}.
For example, the statement \code{y\,=\,x\,+\,2} marks \code{y} as tainted if \code{x} is tainted, as shown in \autoref{fig:taint-propagation}.
\polytaint extends this idea to functions by marking them as tainted if tainted variables are manipulated in their bodies.


\polytaint implements taint propagation via \emph{AST instrumentation}.
This is done by wrapping variable reads/writes, function calls (see \S\ref{sec:ast-instrumentation}), \etc and testing for nodes that access secure values.
For example, in our illustrative example (\autoref{lst:secure-mc}), the variable assignments at lines~\ref{lst:python-regression:m} and~\ref{lst:python-regression:c} will be wrapped and evaluated by \polytaint to obtain the corresponding variables (\ie \code{m} and \code{c}) that receive secure values (foreign objects in \truffle terminology) from secure nodes.
We refer to such variables as \emph{taint sources}. 

\smallskip\noindent\textbf{Program function classification.}
The primary goal of taint propagation in \polytaint is to determine which program functions access secure values (directly or indirectly from the taint sources) at run time, and which functions do not. As such, \polytaint classifies functions into three categories: \emph{trusted}, \emph{untrusted} and \emph{neutral}.

\smallskip\noindent\emph{Trusted functions.}
These are functions that manipulate secure variables explicitly within their bodies \ie instantiate secure variables (\ie taint sources) or modify the values of secure variables.
We define a \emph{secure variable} as a program variable which explicitly receives its value from a \secv node via the polyglot API (\eg \code{m} and \code{c} in \autoref{lst:secure-mc}) or a program variable which gets tainted following \polytaint's taint propagation rules.
As such, function \code{trainModel} in \autoref{lst:python-regression} will be tagged as a trusted function by virtue of the variable write: \code{Y\_pred\,=\,m\,*\,X\,+\,c} which uses secure variables \code{m} and \code{c} inside \code{trainModel}'s body.

Trusted functions are included fully inside the trusted/enclave part of the partitioned application.
These functions have \emph{proxy functions} in the untrusted part.
We define the \emph{proxy function} of a program function \code{f} as a function which has the same signature as \code{f} but whose body is stripped and replaced with an enclave transition (\ie \ecall or \ocall) which transfers execution control to \code{f}.
Following this logic, the proxies of \emph{trusted} functions perform \ecall transitions.
The primary rationale for this design choice is \emph{security}: that is, secure variables that contain sensitive information cannot be accessed in the untrusted side.

\smallskip\noindent\emph{Neutral functions.}
These are functions which access secure variables \emph{only} through their input parameters and do not explicitly access already tainted (\ie not considering the tainted inputs) secure variables within their bodies.
In other words, a neutral function does not access a tainted value when none of its arguments is tainted, and may access tainted values if at least one of its arguments is tainted.
In \autoref{lst:python-regression} for example, \code{arraySum} is considered a neutral function by virtue of the tainted inputs \code{Y\,-\,Y\_pred} and \code{X\,*\,(Y\,-\,Y\_pred)}.
Moreover, untrusted code could also perform array sums without sensitive inputs.

Neutral functions are included fully in both the trusted and untrusted partitions; there are no proxy functions involved. The rationale behind this design choice is twofold: the approach allows for good \emph{security} and \emph{performance}. That is, enclave code can access the functionality of the neutral function without leaking sensitive information, via input parameters, to the outside (security), and the code out of the enclave does not need to perform an expensive \ecall transition to access the functionality of the neutral function (performance). Utility functions such as those that calculate generic sums, products, \etc are examples of neutral functions. 

\smallskip\noindent\emph{Untrusted functions.}
These are functions which do not have any access to secure variables during the lifetime of the program. 
Untrusted functions are included fully only in the untrusted partition.
However, they have \emph{proxies} in the trusted side which perform \ocall transitions to access the functionality in the untrusted side of the partitioned application.
The rationale for this design choice is \emph{TCB size reduction}, and thus \emph{security}, \ie code which is not sensitive need not be in the enclave following the principle of least privilege \cite{saltzer1975protection}.

\subsection{AST instrumentation}
\label{sec:ast-instrumentation}

To identify the trusted ($T$), neutral ($N$) and untrusted ($U$) functions, \polytaint uses a dynamic analysis technique.
It executes the program once during the development phase. During this execution, it identifies the secure values, and then deduces the state of the functions, \ie \emph{trusted}, \emph{untrusted}, and \emph{neutral}.
During the in-vitro execution, \polytaint performs taint propagation by leveraging the \truffle instrumentation framework, which makes it possible to intercept the execution of AST nodes.
To instrument the \truffle AST, \polytaint intercepts the executions of these node types: \emph{variable read/write}, \emph{object field read/write},
\emph{array element read/write} and \emph{function call} nodes (see \S\ref{instrumentation_agents}).
\polytaint provides an event node class for each of the node types to be wrapped, as well as a base event node class \code{PolyTaintNode} which all the event node classes inherit from.
\code{PolyTaintNode} implements common functionality which is used by the event node classes.

\polytaint maintains a map data structure called the \emph{taint map} which tracks program symbols (\eg variables, functions, \etc) that access secure values.
A unique \emph{string identifier} is associated to each program symbol such as \eg a variable or function.
The \emph{taint map} associates \emph{string identifiers} to \emph{taint labels}.
We have 3 \emph{taint labels}: $1$ for tainted/trusted nodes, $2$ for neutral nodes, and $0$ for untrusted nodes (\ie representing untrusted functions).

\begin{algorithm}
	\small
	\caption{--- Pseudo-code to check for tainted nodes in AST}\label{algo:ast-traversal}
	\begin{algorithmic}[1]
		\Procedure{TraverseAST}{\textit{node}}  
		 \If{\textsc{IsTainted}(\textit{node})}
		 	\State \Return \textbf{true}
		 \EndIf
		\For{Node \textit{child} \textbf{in} \textit{node}.\textsc{GetChildren}()}
			\If{\textsc{TraverseAST}(\textit{child})}
				\State \Return \textbf{true}			
			\EndIf
		\EndFor	
			\State \Return \textbf{false}
		\EndProcedure
\vspace{5pt}
		\Procedure{IsTainted}{\textit{node}}
			\State String \textit{id} $\gets$ \textsc{GetIdentifier}(\textit{node})
			\If{\textit{taintMap}.\textsc{Contains}(\textit{id})}
				\State \Return \textit{taintMap}.\textsc{Get}(\textit{id}) = 1
			\Else
				\State boolean \textit{polygotEvalTest} $\gets$ \textit{node} \textbf{is} \textit{polyglot}.\textsc{Eval} call
				\State boolean \textit{secVLiteralTest} $\gets$ \textit{node} \textbf{has} "secV" literal node \textbf{as} input
				\State \Return \textit{polyglotEvalTest} $\wedge$ \textit{secVLiteralTest}
			\EndIf 	
		\EndProcedure
	\end{algorithmic}
\end{algorithm}

During instrumentation, \polytaint leverages the\\ \code{onReturnValue} callback of event nodes to test for tainted program symbols in a node's AST.
This is done by recursively traversing the node's child nodes and checking if the child nodes are associated with tainted symbols.
As illustrated in \autoref{algo:ast-traversal}, \polytaint considers that a child node is tainted if
\emph{(i)}~the child node corresponds to a call to the polyglot API with the SecV language identifier (\ie \emph{"secV"}) or
\emph{(ii)}~the child node is a value that was tainted before.
In the remainder of this section, we outline how \polytaint performs taint tracking for the different intercepted nodes.

\smallskip\noindent\textbf{Variable write}:
This is a statement that assigns a value to a program variable.
For example the \js statement:\\
\code{var\,x\,=\,y\,+\,Polyglot.eval("secV",\,"sInt(4)")} inside a function's scope is considered a local variable assignment.
Similarly, in our illustrative example, statements like\\
\code{m\,=\,polyglot.eval(language="secV",\,string="sDouble(0.0)")},\\
\code{c\,=\,polyglot.eval(language="secV",\,string="sDouble(0.0)")}, and \\
\code{Y\_pred\,=\,m\,*\,X\,+\,c}
are variable assignments.

The \truffle instrumentation API provides a generic standard tag operation \code{StandardTags.WriteVariableTag} \cite{standardtags} to identify nodes corresponding to variable write statements in all \truffle languages.
At run time, \polytaint wraps every variable write node and creates a corresponding \emph{variable write event node}.
The \truffle API makes it possible to obtain a node object descriptor \cite{nodedescriptor} which provides the unique name of the program variable being written to, as well as the corresponding AST node.
In the \code{onReturnValue} callback of the \emph{variable write event node}, \polytaint leverages node object descriptors to obtain the exact names of the target program variables being written to.
For example, for the 3 aforementioned variable write statements, the variable names are respectively \code{m}, \code{c} and \code{Y\_pred}.

\polytaint uses the variable names to construct the unique String identifiers for these variables.
Still in the \code{onReturnValue} callback, \polytaint checks the \emph{taint map} for the presence of \code{m}, \code{c} and \code{Y\_pred}.
If they are present, the \code{onReturnValue} callback returns.
Otherwise, \polytaint leverages the \truffle API to obtain the \emph{right-hand side (rhs)} expression node associated with the variable write statements.
\autoref{algo:ast-traversal} traverses the child nodes of the \emph{rhs} expressions to check for the presence of any tainted nodes.

As explained previously, the presence of child nodes corresponding to the \code{polyglot.eval} call as well as the \code{"secV"} string literal as one of its input arguments tests positive for a \emph{taint source}.
This will mark variables \code{m} and \code{c} as tainted in the \emph{taint map}, \ie taint labels of $1$ for their identifiers.
Similarly, \code{Y\_pred} will be marked as tainted due to the presence of tainted variables \code{m} and \code{c} in the \emph{rhs} expression involved in the assignment of \code{Y\_pred}.
If a tainted variable is written in a function, this function is also tagged as tainted in the \emph{taint map}.
For example, the variable write node corresponding to the statement \code{Y\_pred\,=\,m\,*\,X\,+\,c} is tainted as a result of the presence of tainted variables \code{m} and \code{c} in the statement, therefore the enclosing function $trainModel$ is marked as tainted in the \emph{taint map}.
\polytaint calls \code{getName} on the root node object \cite{trufflenode} to obtain the unique name of the enclosing function/method for any instrumented node. 

\smallskip\noindent\textbf{Variable read}:
This is a statement that reads the value of a program variable. Variable read operations could occur in an if/else statement, a for loop, a function call via parameters, \etc.
The \truffle instrumentation API provides a generic standard tag \\\code{StandardTags.ReadVariableTag} \cite{standardtags} to identify nodes that correspond to variable read statements in all \truffle languages.

At run time, \polytaint wraps every variable read node and creates a corresponding \emph{variable read event node}.
The event node's \code{onReturnValue} callback is used to obtain the unique name of the variable being read from the node object descriptor, and the \emph{taint map} is checked for the presence of the variable's identifier.
If a tainted variable is read in a function (\eg \code{while(i\,<\,tainted\_variable))}, this function is also tagged as tainted in the \emph{taint map}.
If a tainted function argument is read (\eg reading the value of \code{Y\_pred} in \code{arraySum(Y\,-\,Y\_pred)} on line~\ref{lst:python-regression:array}), the function node is tracked as a neutral function in the \emph{taint map} (\ie taint label = $2$) if it has not been tagged as tainted.
In other words, a taint label of $1$ for tainted/secure function is preferred for the function node over a taint value of $2$ for neutral.
As such, \code{arraySum} is tagged as a neutral function in the \emph{taint map} .

\smallskip\noindent\textbf{Object field write:}
This operation assigns a value to an object's field or property.
At the time of this writing, the \truffle API does not provide standard tags to identify generic object field write nodes but \truffle languages typically provide language-specific tags to identify such nodes.
For example, \truffle-\js provides the \code{WritePropertyTag} \cite{jstags} to identify object property writes.
In some languages like \js, the global scope is an object and global variables are properties of this object.
As such, global variable writes in \js are instrumented via the \code{WritePropertyTag}. Object field writes are instrumented with \emph{object field write event nodes} in a similar fashion to variable writes.


\smallskip\noindent\textbf{Object field read:}
This operation reads an object's field or property.
Similarly to object field writes, since there are no generic standard tags yet in the \truffle API to identify object field read nodes, \truffle languages typically provide language-specific tags like \code{ReadPropertyTag} \cite{jstags} in \truffle-\js to identify an object field read node.
Object field reads are instrumented with \emph{object field read event nodes} in a similar fashion to variable reads.

\smallskip\noindent\textbf{Array element write (resp. read):}
This is an operation that writes (resp.~reads) a value to an array object,\\ \eg \code{array[0]\,=\,4} (resp.~\code{if(array[3]\,<\,0)}).
Array element writes\\(resp.~reads) are instrumented in a similar fashion to object field writes (resp.~reads), with \emph{array element write} (resp.~\emph{read}) \emph{event nodes}.


\smallskip\noindent\textbf{Function call:}
This is an expression that passes control (and possible arguments) to a function or method in a program.
The \truffle instrumentation API provides a generic standard tag \\\code{StandardTags.CallTag} \cite{standardtags} to identify language nodes that correspond to guest language functions and methods in all \truffle languages.
At run time, \polytaint wraps every function call node and creates a corresponding \emph{call event node}.
The call event node's \code{onReturnValue} callback is used to test first for tainted arguments.

\truffle languages usually provide methods to obtain argument nodes during instrumentation, \eg \code{getArgumentNodes} provided by \truffle-\js (\code{JSFunctionCallNode} \cite{jsfuncnode}) and \truffle-\python\\ (\code{PythonCallNode} \cite{pythoncallnode}).
If an argument node for a function corresponds to a tainted symbol, \eg a tainted variable, this function is tagged as neutral in the \emph{taint map} if it has not been tagged as tainted.
For example, \code{arraySum} is tagged as neutral by virtue of the presence of the tainted variable \code{Y\_pred} in the argument expression.

\polytaint maintains a list (\emph{seen list}) that comprises all functions/methods that have been visited during execution at run time.
Every function/method in the list is identified as a \code{PolyTaintFunction} object, which is an instance of \code{PolyTaintFunction} class.
This class defines attributes representing the actual function node (\ie \code{Node} object), a list representing the different argument types passed to the function (\eg \code{int} for \code{trainModel}, \code{Object} for \code{arraySum}, \etc), a \code{String} representing the return type (\eg \code{double} for \code{arraySum}, \code{void} for \code{trainModel}, \etc) of the function, and an integer value representing the function's taint label.
The \code{onReturnValue} callback of every event node contains a \code{VirtualFrame} \cite{truffleframe} parameter comprising the actual argument values passed to the instrumented node, \eg a function call node, and an \code{Object} parameter representing the actual result received after the node is executed.
\polytaint leverages these two parameters to obtain the corresponding argument types passed (if they exist) to every function call node and return types, respectively.
The input and return types are used later on during program partitioning (\S\ref{sec:partitioning}).

At the end of program instrumentation with \polytaint, all application functions which have been executed at run time (in the \emph{seen list}), but have not been tagged as trusted/tainted or neutral in the \emph{taint map} are considered untrusted functions.
As such, for our illustrative example, after instrumentation, we should have set $T$: \code{trainModel}, set $N$: \code{arraySum}, and set $U$: \code{readXData} and \code{readYData}.
This information is then passed to the program partitioner which builds two programs representing the trusted and untrusted partitions of the instrumented program.

\subsection{Program partitioning}
\label{sec:partitioning}

The aim of the partitioning stage is to separate the original program into two parts: a \emph{trusted} part which executes inside the enclave and comprises functions $T\cup N$, and an \emph{untrusted} part which executes outside the enclave and comprises $U\cup N$.
By removing the $U$ functions from the enclave, we reduce the size of the TCB. Furthermore, eliminating $U$ functions also eliminates any functions of the language runtime's system libraries invoked by $U$, which are often very large (as stated in the introduction).


In order to interpret the ASTs of the $N$ and $T$ functions inside the enclave at runtime, we have to also embed the corresponding \truffle language interpreter in the enclave. However, \truffle is written in Java, which means that we need a Java runtime inside the enclave to execute the \truffle interpreter.
Since embedding a full JVM with its system library inside the enclave would increase the size and the attack surface of the TCB to an unacceptable degree, we choose to base the design of our partitioning tool on native images (\S\ref{native-image}).
\graal's Native Image technology makes it possible to include only reachable program elements (\ie methods, classes, and objects) into the resulting native image, making it suitable for restricted environments like Intel SGX enclaves.
Any unused classes in the \truffle interpreters or \graal's runtime components (\eg garbage collector) are pruned out of the native images.
The remainder of this section describes our partitioning approach.


\subsubsection{Generated functions}
\graal AoT only supports compiling Java applications to create native images (\ie no \js, \python, \etc), therefore, we need to find a way to run the partitioned guest application code (\eg \js, \python, \etc) via native images.
To achieve this, the \emph{partitioner} leverages \truffle's polyglot API to embed the guest code inside two Java programs: \code{Trusted.java} for the trusted partition and \code{Untrusted.java} for the untrusted partition.
This is done by creating static Java methods that correspond to each of the methods seen during taint analysis (\ie $T$, $N$, and $U$).
These static methods simply execute the corresponding ASTs of the guest functions they represent.

For instance, \autoref{lst:js-multi} illustrates how the AST of a \js function \code{multi} can be executed from within a Java application, and how a Java method \code{hello} can be executed from within a \js program.


\begin{lstlisting}[language=Java,
xleftmargin=3pt,
xrightmargin=25pt,
label={lst:js-multi},
caption={Java host and guest \js interaction via polyglot API}.]
import org.graalvm.polyglot.*;
public class Example {
	public static void main(String[] args) {
		Context ctx = Context.newBuilder().allowAllAccess(true).build();
		Value jsMulti = ctx.eval("js", "(function multi(a, b){return a*b;})");
		Value res = jsMulti.execute(6, 7);
		System.out.println(res); // prints 42 in Java scope
		Value jaHello =
      ctx.asValue(Polyglot.class).getMember("static").getMember("hello");
    String jsStringFunc = "function jsHello(jaHello){jaHello();}jsHello;"
		Value jsFunction = ctx.eval("js", jsStringFunc);
		jsFunction.execute(jaHello); // prints "Hello Java" in js scope
	}
	public static void hello() {
		System.out.println("Hello Java");
	}	
}
\end{lstlisting}

Using the same idea outlined in \autoref{lst:js-multi}, \code{Trusted.java} therefore comprises static Java methods for \texttt{\\trainModel}, \code{arraySum}, \code{readXData} and \code{readYData}.
The static Java methods for \code{trainModel} and \code{arraySum} execute the corresponding AST code, \ie the actual guest source code for those functions, while the static Java methods for \code{readXData} and \code{readYData} (proxy methods) perform \ocall transitions to execute the real methods in the untrusted partition (see below).
The actual source code of a function is obtained via the \code{getSourceSection} method \cite{trufflenode} of the \truffle \code{Node} class.
This is illustrated by \autoref{lst:trusted}. 

\begin{lstlisting}[language=Java,
xleftmargin=15pt,
xrightmargin=25pt,
caption={Trusted.java after partitioning},label={lst:trusted}]
public class Trusted {	
    Context ctx = Context.newBuilder().allowAllAccess(true).build();
    public static void trainModel(int iter){
        Value arraySumVal = ctx.asValue(Trusted.class).
        		getMember("static").getMember("arraySum");
        ctx.eval("python", "//trainModel code").
        		execute(iter, arraySumVal, ...);
	 }
 	public static double arraySum(Object array){
 		double ret = ctx.eval("python", "// arraySum code here").
 				execute(array).asDouble();
 		return ret;	
	 }
 	public static Object readXData(int n){
 		ocall_readXData(n);
 	}
 	...
}
\end{lstlisting}

 Similarly, \code{Untrusted.java} comprises static Java methods for \code{trainModel}, \code{arraySum}, \code{readXData} and \code{readYData}.
 The static Java methods for \code{readXData}, \code{readYData} and \code{arraySum} execute the corresponding AST code for those functions while the static Java method for \code{trainModel} performs an \ecall transition to execute the method in the trusted partition.
\autoref{lst:untrusted} exemplifies this.
 
 \begin{lstlisting}[language=Java,
xleftmargin=5pt,
xrightmargin=20pt,
caption={Untrusted.java after partitioning},label={lst:untrusted}]	
public class Untrusted {
	Context ctx = Context.newBuilder().allowAllAccess(true).build();
 	public static Object readXData(int n) {
 		return ctx.eval("python","// readXData code here").execute(n);			
 	}
 	public static void trainModel(int n) {
 		ecall_trainModel(n);		
 	}
 	...
}
\end{lstlisting}

\subsubsection{Transition between the partitions}
With our program partitioned, we must allow communication between the two partitions. That is, how does a Java method in \sloppy{\code{Trusted.java}} (\ie the trusted partition) invoke another Java method in \code{Untrusted.java} (\ie the untrusted partition)?

To achieve this, we leverage \graal \emph{C entry points} \cite{centrypoint,montsalvat}.
These are special Java methods in a native image program which are callable from a regular C/C++ program or another native image.
As such, C entry point methods can act as relays between the trusted and untrusted partitions.
That is, a proxy function in partition-$x$ can invoke the corresponding real static method (\code{m}) in the opposite partition-$y$ if there is a \emph{C entry point} method for \code{m} in partition-$y$.

For the trusted partition (\ie \code{Trusted.java}), the program partitioner generates C entry point methods corresponding to all the trusted functions.
These entry point methods are the target of \ecalls from the untrusted code calling a trusted function in the final SGX application.
Similarly, for the untrusted partition (\ie \code{Untrusted.java}), C entry point methods are generated for the untrusted functions.
These entry point methods are the target of \ocalls from the trusted partition calling an untrusted function. 

\subsubsection{Serialisation}
\label{sec:serialisation}

Because objects cannot be sent across an enclave boundary \cite{sgxsdk,montsalvat}, any static methods that have non-primitive input or return types (\eg arrays, strings, \etc) to be transferred across the enclave serialise the input or return values into a byte array.
The byte array is then marshalled across the enclave boundary in the corresponding \ecall or \ocall, and deserialised on the opposite side into a Java \code{Object}.
Serialisation and deserialisation are done using Java's \code{ObjectOutputStream} and \code{ObjectInputStream} \cite{object-streams} classes, respectively.

\subsubsection{Execution of a function} 
Thanks to native image, the trusted and the untrusted partitions execute the (binary) compiled version of the corresponding \truffle guest language interpreter at runtime.
At runtime, the guest language interpreter parses the string that represents the guest code, builds the AST, and executes it. When the AST of a function \code{f} contains a call to another function \code{g}, the symbol table of the interpreter may not contain the symbol \code{g}. In that case, the \truffle API leverages Java reflection to find a static Java method that corresponds to the symbol \code{g} and executes it. As an example, in \autoref{lst:js-multi}, Java reflection is used to expose the Java function \code{javaHello} (or the JS function \code{multi}) to \truffle interpreter when executing \code{jsHello}.


\subsection{Building native images and the SGX program}
\label{sec:ImageGenerator}

The aim of ImageGenerator is to AoT compile \code{Trusted.java} and \code{Untrusted.java} into relocatable object files (\code{trusted.o} and \code{untrusted.o} respectively).
When building a native image, \graal makes it possible to provide which guest language implementations (\ie the \truffle interpreters) should be made available in the resulting native image.
In the case of \sys, all guest languages involved in the partitioned program are provided as inputs to the \emph{native image generator}.
This tool performs static analysis (see \S\ref{native-image}) \cite{nativeImgs} to determine the reachable Java elements, \ie the classes, methods, and objects from the Java program that is being compiled and the \truffle language implementations that are required to run the corresponding Java programs.
These reachable components are then AoT compiled into native images: \code{trusted.o} and \code{untrusted.o}.

It is worth noting that the native image builder does not AoT compile guest language code (\eg \js, \python, \etc) that is embedded in the Java programs.
Indeed, the guest language code will be interpreted or JIT compiled by \graal at run time.
The \ecall and \ocall definitions are compiled into object files and linked with \code{trusted.o} and \code{untrusted.o}, as well as SGX C/C++ library code, to create the final Intel SGX application.
A small shim library is included inside the enclave which seamlessly relays unsupported system calls (\eg \code{read}, \code{write}, \etc) to the untrusted runtime via \ocalls, which perform the real system calls and return the results back to the enclave.

\section{Evaluation}
\label{sec:evaluation}

The experimental evaluation of \sys seeks to answer the following research questions:

\begin{itemize}[leftmargin=28pt,nosep]
	\item[\textbf{RQ1}:] What is the implementation effort for a multi-language tool (\ie \sys) as compared to language-specific tools? (\S\ref{sec:implem-effort})
	
	\item[\textbf{RQ2}:] What is the cost of injecting \sys nodes into a program AST? (\S\ref{sec:cost-of-secv-nodes})
	\item[\textbf{RQ3}:] What is the cost of taint tracking with \polytaint? (\S\ref{sec:taint-tracking-cost})
	\item[\textbf{RQ4}:] How does partitioning affect application performance? (\S\ref{sec:partitioning-impact})
\end{itemize}

\subsection{Experimental setup}
\looseness-1
Our evaluation is conducted on a server equipped with a quad-core Intel Xeon E3-1270 CPU clocked at 3.80\,GHz, and 64\,GB of DRAM. 
The processor has 32\,KB L1 instruction and data caches, a 256\,KB L2 cache, and a 8\,MB L3 cache. 
The server runs Ubuntu 18.04.1 LTS 64\,bit and Linux 4.15.0-142. 
We run the Intel SGX platform software with version 2.16 of the SDK and and version 2.14 of the driver.
The EPC size on this server is 128\,MB, of which 93.5\,MB is usable by enclaves.
The enclaves have maximum heap sizes of 8\,GB and stack sizes of 8\,MB.
All native images are built with a maximum heap size of 4\,GB.
We use \graal version 22.1.0. 
All reported measurements are averaged over 5 runs.

\begin{table}[!t]
	\setlength{\tabcolsep}{2.5pt}
	\resizebox{\columnwidth}{!}{
	\begin{tabular}{l c l r}
		\toprule
		\textbf{Partitioning tool} & \textbf{Supported languages} & \textbf{Frameworks used + LoC}&  \textbf{Tool's LoC}\\
		\midrule
		Civet & Java & SOOT + Phosphor ($\approx 421K$ LoC) & 6,870\\
		Montsalvat  & Java & Javassist ($\approx 38K$ LoC) & 3,500\\
		Glamdring & C & Frama-C + Program slicer ($\approx 90K$ LoC)  &5,000\\		
		\sys & \js, \python, R, \etc &Truffle ($\approx 276K$ LoC)& 4,880\\
		\bottomrule\\
	\end{tabular}
}
	\caption{Total lines of code for existing partitioning tools and their underlying frameworks.}
 \label{tab:implem-effort}
 \vspace{\imagespace}
\end{table}


\subsubsection{\textbf{Implementation efforts (RQ1)}}\emph{What is the implementation effort for a multi-language tool as compared to language-specific tools?}
\label{sec:implem-effort}
\autoref{tab:implem-effort} reports the total lines of code (LoC) of various enclave code partitioning tools, as well as the LoC for the underlying frameworks used by these tools.
We observe that \sys's code base is $\approx 1.41\times$ and $\approx 1.02\times$ smaller as compared to the full code bases of Civet and Glamdring, respectively. We argue that the \sys approach is better in terms of code simplicity (leverages a single self-contained framework) and efficiency (achieves the desired functionality with fewer lines of code). Furthermore, \sys provides a single extensible multi-language system in relatively fewer LoC, contrary to the other tools which are language specific.

\subsubsection{\textbf{Overhead of SecV nodes (RQ2)}}\emph{What is the cost of injecting \sys nodes into a program's AST?}
\label{sec:cost-of-secv-nodes}
\begin{figure}[!t]
	\centering
	\includegraphics[scale=0.67]{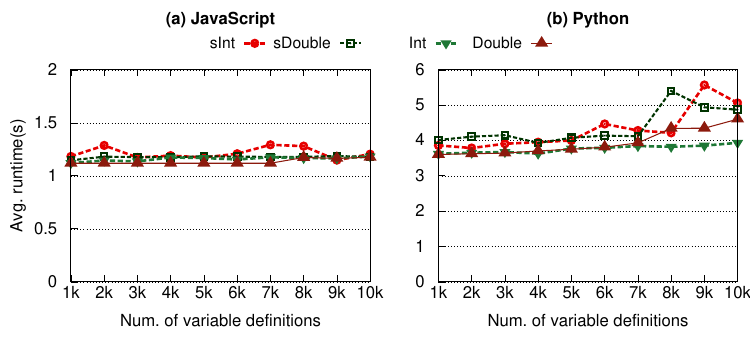}
	\caption{Cost of using secure values in \js and \python programs.}
	\label{fig:secl-type-cost}
	\vspace{\imagespace}
\end{figure}

The goal of this experiment is to measure the overhead caused by the use of secure nodes in a polyglot program. In this regard, we generate synthetic programs in \js and \python that comprise functions with varying numbers of variables that receive secure values (\eg secure \code{int}, \code{double} values) via the polyglot API. We compare the run time of the functions with a similar setup where the functions define regular values, \ie no injection of \secv nodes via the polyglot API.
\autoref{fig:secl-type-cost} shows the results obtained for different value types used.


\smallskip\noindent\emph{Observation.}
For \js programs, using \code{sInt} variables is $\approx 1.05 \times$ slower on average as compared to using regular integer variables without the polyglot API.
Moreover, using \code{sDouble} variables is $\approx 1.04 \times$ slower on average as compared to using regular \code{double} type variables without the polyglot API.
For \python programs, using \code{sInt} variables is $\approx 1.14 \times$ slower on average as compared to using regular \code{int} type variables without the polyglot API.
Furthermore, using \code{sDouble} variables in \python is $\approx 1.11 \times$ slower on average as compared to using regular \code{double} type variables.

\smallskip\noindent\emph{Discussion.}
For both \js and \python, using \secv nodes involves calls to the \truffle polyglot API, as opposed to defining regular program variables for which there is no intermediate API involved.
This explains the additional cost when introducing secure values in the program. In practice, programs will typically define few secure variables leading to a small overhead relative to the total runtime cost of the full program


\observ{
The cost of injecting secure nodes into a program's AST is small.
}

\subsubsection{\textbf{Overhead of taint tracking (RQ3)}}\emph{What is the cost of taint tracking with \polytaint?}
\label{sec:taint-tracking-cost}
\begin{figure}[!t]
	\centering
	\includegraphics[scale=0.67]{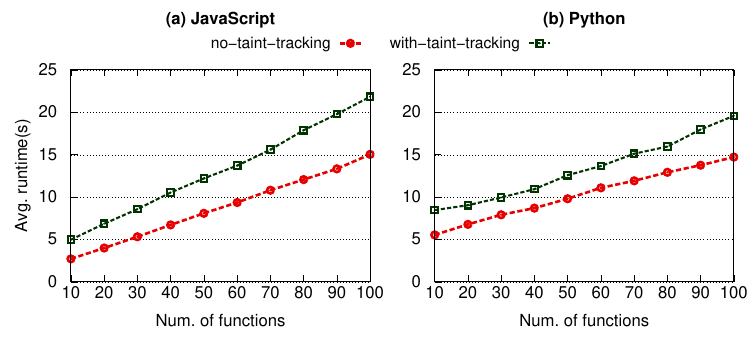}
	\caption{Cost of taint tracking with \polytaint for \js and \python.}
	\label{fig:taint-track-cost}
	\vspace{\imagespace}
\end{figure}

In this experiment, we aim to measure the performance of taint tracking with \polytaint.
To that end, we generate synthetic polyglot programs in \js and \python.
The synthetic programs consist of a varying number of functions that all perform a bubble sort on a \secv array of 1{,}000 randomly generated numbers.
We chose a bubble sort algorithm here because it performs many variable read/write operations, which trigger the creation of taint tracking instrumentation nodes in \polytaint.
The programs run on \graal using their corresponding \truffle interpreters, with or without \polytaint taint tracking enabled.
\autoref{fig:taint-track-cost} shows the comparative performance of both setups.

\smallskip\noindent\emph{Observation.}
The experimental results show that the \js program running with \polytaint taint tracking enabled is $\approx 1.56\times$ slower when compared to the \js program running without taint tracking enabled. Similarly, for the \python program running with \polytaint taint tracking enabled, it is $\approx 1.31\times$ slower on average when compared to the variant without taint tracking enabled.

\smallskip\noindent\emph{Discussion.}
For both \js and \python programs, \polytaint introduces more nodes (\ie wrappers and execution event nodes) to their ASTs during instrumentation.
As seen in \S\ref{sec:ast-instrumentation}, those nodes perform operations to track tainted variables, which explains the performance decrease (\ie longer runtime) when taint tracking is enabled, as compared to the variants where the programs run without any taint tracking.

\observ{
Taint tracking via AST instrumentation introduces overhead at run time.
This overhead is due to the additional operations performed by instrumentation nodes introduced in the program's AST.
}


\subsubsection{\textbf{Effect of partitioning on performance (RQ4)}}\emph{How does partitioning affect application performance?} 
\label{sec:partitioning-impact}

To study the performance impact of partitioning on application performance with \sys, we first use a synthetic polyglot benchmark written in \js.
We then partition two real-world applications that implement the \emph{PageRank} algorithm \cite{pagerank} and the \emph{linear regression} ML algorithm \cite{regressionXu2022}.

\begin{figure}[!t]
	\centering
	\includegraphics[scale=0.745]{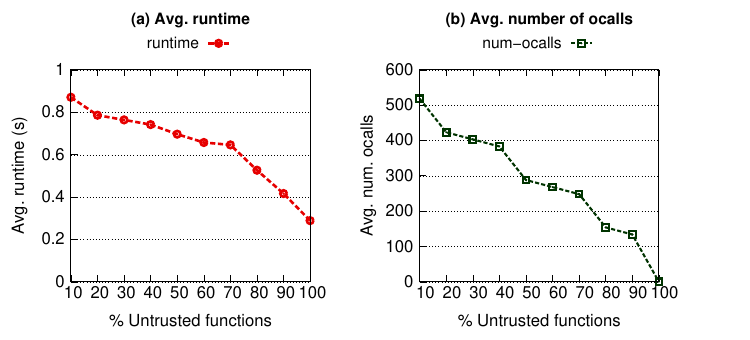}
	\caption{Effect of program partitioning on a generic synthetic program.}
	\label{fig:micro-synthetic}
	\vspace{\imagespace}
\end{figure}

\smallskip\noindent\textbf{Synthetic benchmark.} 
Our synthetic benchmark is a \js program that comprises 100 functions that each perform a bubble sort on a local array variable of size 100.
We leverage secure values (secure array or regular array) in a varying percentage of the functions, analyse the polyglot programs with \polytaint and partition these programs for enclaves.
The purpose of our synthetic benchmark is to highlight the effect of partitioning on a generic program.
\autoref{fig:micro-synthetic} shows the results.

We observe that as more functions are partitioned out of the enclave, the overall performance of the program improves.
This is explained firstly by the reduced number of \code{libc}-related \ocalls performed by the embedded  \graal runtime components (language interpreters, GC, \etc), and secondly by the fact that we have less overhead due to enclave data encryption-decryption operations to and from the EPC by the MEE \cite{montsalvat}.

\observ{
Enclave performance improves as more computations are delegated to the untrusted side.
This is due to less expensive enclave context switches, \eg via ocalls, as well as less expensive enclave-related cryptographic operations.
}

We have observed the effect of partitioning on a generic synthetic program.
We now leverage \secv to partition real-world applications.
For these applications, we evaluate 3 modes for running the programs: entirely inside the enclave (\code{no-part}), as a normal polyglot native image without SGX (\code{native}), and as a partitioned native image using \sys (\code{part}).
The partitioning scheme adopted for each application is not necessarily realistic, but aims to highlight and explain the performance improvement obtained after partitioning.


\begin{figure}[!t]
	\centering
	\includegraphics[scale=0.745]{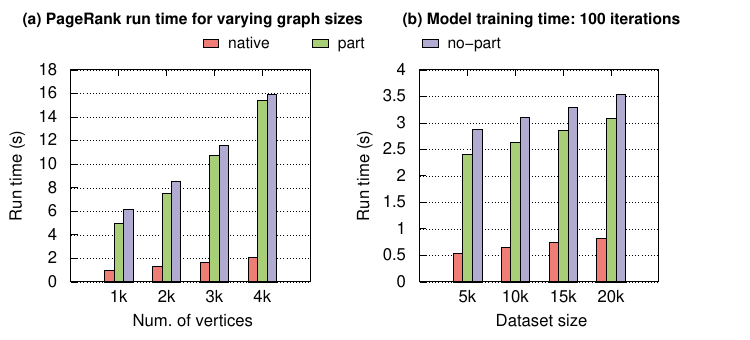}
	\caption{Results for partitioning the PageRank and Linear Regression programs.}
	\label{fig:macro-all}
	\vspace{\imagespace}
\end{figure}

\smallskip\noindent\textbf{PageRank.}
PageRank \cite{pagerank} is a popular algorithm that is used in graph processing frameworks \cite{graphchi} to weigh the relative importance of nodes (\ie node ranks) in a directed graph.
One could envision a scenario where the node ranks are to be secured within an enclave.
We leverage \sys to partition a PageRank program with the node rank data structure (a secure array) tagged as a secure variable. The PageRank program comprises a function to generate a directed graph and obtain the corresponding adjacency list (graph pre-processing), and other functions which use the adjacency list to perform the PageRank algorithm.
The graphs are generated using the RMAT algorithm \cite{rmat}, and all graphs with $n$ vertices have $2\cdot n$ edges.
\autoref{fig:macro-all} (a) shows the results obtained when running partitioned and unpartitioned versions of PageRank with varying graph sizes.

\smallskip\noindent\emph{Observation.}
The partitioned version of the PageRank program is $\approx 1.12\times$ faster on average (\ie about $10.8\%$ performance improvement) as compared to the unpartitioned version.
Furthermore, we have on average $\approx 2.53\times$ fewer \ocalls in the partitioned program as compared to the unpartitioned one.
The native version of the program (\ie no SGX) is $\approx 6.85\times$ and $6.17\times$ faster on average as compared to the unpartitioned and partitioned versions, respectively. 

\smallskip\noindent\emph{Discussion.}
After partitioning the program, the functions responsible for graph generation and pre-processing are moved to the untrusted partition, while those that perform the PageRank algorithm to obtain node ranks remain in the trusted partition.
At runtime, the pre-processed graph (in the form of an adjacency list) is serialised in the untrusted partition and marshalled into the enclave (see \S\ref{sec:serialisation}) via an \ocall.
It is then deserialised to recreate the adjacency list object which is used by the in-enclave PageRank algorithm to compute the final page ranks.
The computations outside the enclave runtime do not incur expensive context switches via \ocalls, or SGX-related cryptographic operations.
That explains the performance improvement as well as the reduced number of \ocalls in the partitioned version as compared to running the full program inside the enclave.
This also explains why the native application is fastest: no SGX-related cryptographic operations, no enclave context switches, and less memory restrictions (the enclave has only about $93MB$ of memory). However, it is the least secure of all.

\smallskip\noindent\textbf{Linear regression.}
Securing ML models with TEEs is common \cite{plinius,secureml}. We aim to partition a linear
regression program for an enclave runtime.
Similar to PageRank, a linear regression program typically consists of functions to read and pre-process (\ie normalisation and standardisation) the input datasets to be used, and other functions which use the dataset to train the ML model.
We leverage \sys to specify secure variables (\ie \code{m} and \code{c} in \autoref{lst:python-regression}), analyse and partition the linear regression program.
\autoref{fig:macro-all} (b) shows the results when the linear regression model is trained for 100 training iterations for partitioned and unpartitioned versions of the program.

\smallskip\noindent\emph{Observation.}
The partitioned version of the linear regression program is $\approx~1.17\times$ faster on average (\ie about $14.5\%$ performance improvement) as compared to the unpartitioned version.
Moreover, we have on average $\approx~1.36\times$ fewer \ocalls in the partitioned program when compared to the unpartitioned one.
The \code{native} version of the program is $\approx 4.73\times$ and $4.04\times$ faster on average when compared to the unpartitioned and partitioned versions, respectively. 

\smallskip\noindent\emph{Discussion.}
After program partitioning, the functions for data generation and pre-processing (which do not access \code{m} and \code{c}) are partitioned out of the enclave, while the trusted functions that train the model (and hence access the secure values \code{m} and \code{c}) are part of the enclave partition.
At runtime, dataset pre-processing is done in the untrusted partition; this entails standardising both the \code{X} and \code{Y} datasets.
The standardised datasets (arrays) are then serialised and marshalled into the enclave runtime via \ocalls, deserialised inside the enclave, and used to train the linear regression model.
By offloading the data generation and pre-processing phases to the untrusted runtime, the enclave is relieved of expensive computations, which leads to an overall improvement in application performance when compared to running the full program inside the enclave.
Similar to PageRank, partitioning introduces overhead for data serialisation and deserialisation, but this overhead is offset by the performance gain from performing some computations outside (\ie no expensive \ocalls), rather than inside the enclave.
\observ{
The TCB size of programs can be decreased without degrading performance.
As a matter of fact, the peformance analysis of both PageRank and linear regression shows that decreasing the TCB size via partitioning can lead to better performance: security and performance are improved at the same time.
}

\section{Limitations of our design}
\label{sec:limitations}
\smallskip\noindent\textbf{\textit{Limited code coverage}.~}
The present design of \sys is based on a purely dynamic program analysis approach, where programs are run once during the development phase to deduce the set of \textit{trusted} ($T$), \textit{neutral} ($N$), and \textit{untrusted} ($U$) functions, which are then used to create the trusted and untrusted partitions.
However, this design provides limited code coverage, potentially leaving security-sensitive code out of the enclave; limited code coverage is a fundamental limitation of most dynamic analysis tools \cite{pkru2022}. To address this problem, \textit{static analysis} can be applied, but it has major drawbacks. Firstly, it performs an over-approximation of the code to be tainted (due to polymorphism), which leads to a larger TCB. Secondly, Truffle ASTs exist only at runtime, making static analysis infeasible. Ultimately, a mixture of both static and dynamic taint analysis techniques could be a reasonable compromise. We leave this as future work.

\smallskip\noindent\textbf{\textit{Application termination}.}
During analysis, different program runs (with varying inputs) could result in different execution paths being taken at runtime. This could in turn result in different (and possibly conflicting) sets for $T$, $N$, and $U$. From a security perspective, such a scenario may lead to secure values being leaked to the untrusted partition at runtime. In our prototype implementation, we avoid this problem by checking if a value is secure just before serializing it in in-enclave proxies.
Thanks to this check, in the worst case, the application will terminate with an error, but will not let a secure value escape the enclave.
Running multiple tests or using symbolic execution could decrease the probability of terminating the program.
Similarly, we leave this implementation as future work.

\section{Leveraging \sys for other use cases}
\label{sec:discussions}
\sys's design can be extended to encompass more generic compartmentalization of applications to not only enhance security, but also promote modularity and maintainability. In terms of security, the current design can be extended to support Intel's new TEE technology: \textit{trust domain extensions} (TDX)~\cite{cheng2023intel}. This entails running the trusted and untrusted partitions in separate VMs, or “trust domains”. 

In a context unrelated to security, \sys can also be extended to partition applications into smaller loosely coupled microservices that run in dedicated VMs, hence promoting flexibility and maintainability.

\section{Related work}
\label{sec:rw}


We classify related work into 3 categories:
\emph{(i)}~tools that run full, unmodified applications inside enclaves,
\emph{(ii)}~taint tracking tools, and
\emph{(iii)}~code partitioning tools.

\looseness-1
\smallskip\noindent\textbf{Running full applications inside enclaves.}
Various tools like SCONE \cite{scone}, Graphene-SGX \cite{graphene}, TWINE \cite{twine}, and SGX-LKL \cite{sgxlkl} propose solutions to run entire legacy applications inside enclaves. 
Their approach severely increases the size of the TCB, at the risk of added security vulnerabilities. 
\sys provides a generic solution to partition applications for enclaves while trying to keep the TCB as small as possible.

\smallskip\noindent\textbf{Taint tracking tools.}
Several tools exist for taint tracking. 
Phosphor \cite{phosphor} is a dynamic taint tracking tool for Java programs, while Dytan \cite{dytan} performs taint tracking in x86 binaries. 
TAJ \cite{taj} provides efficient static taint analysis for Java applications.
\cite{staicu2020} provide a tool to extract taint specifications for JavaScript libraries.
However, these tools are language-specific (\ie only for Java, only for x86 assembly code, \js, \etc) and cannot be readily leveraged to analyse code in different languages. 
\sys bridges this gap with a language-agnostic taint tracking approach. 
TruffleTaint \cite{kreindl20} leverages the \truffle framework to provide a language-agnostic platform to build dynamic taint analysis applications.
While we leverage very similar instrumentation techniques, we provide a novel way to specify sensitive values through the introduction of secure AST nodes, and leverage these to partition code for the enclave runtime in a language-agnostic fashion.


\smallskip\noindent\textbf{Language specific partitioning tools.}
Several tools have been proposed to partition code written in specific languages for enclaves.
Glamdring \cite{glamdring} provides a technique to automatically partition C applications, while Montsalvat \cite{montsalvat}, Civet \cite{civet}, and Uranus \cite{uranus} propose solutions to partition Java applications for Intel SGX enclaves. 
Among those partitioning systems, none provide a language-independent way to partition enclave programs. \sys solves this problem by introducing a multi-language technique to partition programs for enclaves.

\section{Conclusion and Future Work}
\label{sec:conclusion}
In this paper, we presented \sys, a multi-language approach to analyse and partition programs for Intel SGX enclaves.
\sys provides generic \emph{secure nodes} that encapsulate sensitive program data, and that can be injected into the ASTs of programs written in a wide range of programming languages.
\sys provides a dynamic taint tracking tool, \polytaint, which tracks the flow of sensitive data from \emph{secure nodes} in a program at run time, and partitions the program into trusted and untrusted parts which are executed in and out of the secure enclave, respectively.
Our evaluation of \sys shows it can reduce the program's TCB size without any performance degradation.

\smallskip\noindent\textbf{Future work.~}We plan to extend \sys along two directions:
	
\smallskip\noindent\textit{More data types.~}This entails extending \sys's \emph{secure node generator} to cover more data types and data structures, \eg maps, lists, \etc, as described in \S\ref{sec:new-types}.
	
\smallskip\noindent\textit{Support for more languages.~}\polytaint can be extended to fully cover more \truffle languages like TruffleRuby, FastR, \etc. Regarding LLVM-based languages like C/C++, the Truffle framework provides an LLVM interpreter, Sulong~\cite{sulong}. This interpreter can also be extended to run programs in Go, by leveraging tools like GoLLVM~\cite{gollvm}. Alternatively, Truffle's WebAssembly (Wasm) implementation can be leveraged to provide a common compilation target for supported languages: Rust, Go, Kotlin, \etc. Adding support for these languages in \sys mainly involves incorporating wrapper nodes to handle language-specific semantic constructs and AST nodes in the new language. There is in-depth documentation online to facilitate these extensions.
	

\begin{acks}	
	This work has been supported in part by Oracle donation CR 3801 and project 200021\_178822 of the Swiss National Science Foundation (FNS). We are grateful to the anonymous reviewers and our shepherd, Alexios Voulimeneas, for their valuable feedback.	
\end{acks}




%

\bibliographystyle{plain}
\interlinepenalty=10000
\bibliography{base}

\end{document}